\documentclass[prd,showkeys,twocolumn,amsmath,amssymb]{revtex4}
\usepackage{graphicx}
\usepackage{subfigure}
\usepackage{bm}
\usepackage[colorlinks,citecolor=blue,anchorcolor=red,menucolor=red,linkcolor=red,filecolor=red,runcolor=red,urlcolor=blue,frenchlinks=red]{hyperref}

\begin{document}
%=====================================================================================
\title{Several conjectures from the hadron physics: The transient worlds beside(s) ours}
%=====================================================================================
%=====================================================================================
%

\author{Hua-Xing Chen}
\email{hxchen@seu.edu.cn}
\affiliation{School of Physics, Southeast University, Nanjing 210094, China}

\begin{abstract}
The past decades witnessed the golden era of hadron physics, which gives us a good opportunity to study the physics happening in a transient period of time. The development on the singly heavy baryons indicates that there exists the fine structure of hadron spectrum caused by the direct strong interaction, and the development on the exotic hadrons indicates that the residue strong interaction is capable of forming the hadronic molecules. Similar to the electromagnetic interaction, the strong interaction may be capable of forming some imaginable hadronic worlds. Moreover, there can be various worlds formed by various fundamental physical laws. Some hadrons have so transient lifetimes that they may not even be formed. We discuss whether these non-existent particles are capable of affecting our realistic world. We discuss what kinds of particles exist, and so can be observed, in our realistic world. We conjecture that the ratios $R \equiv M/\Gamma \gg 1/2$ ($M \tau \gg \hbar/2$) and $R \equiv M/\Gamma < 1/2$ ($M \tau < \hbar/2$) can be used to describe the particles existing and not-existing in our realistic world, respectively. Here $M$, $\Gamma$, and $\tau$ are the mass, width, and lifetime, respectively. We propose to use the ratio $R \equiv M/\Gamma \sim 1/2$ ($M \tau \sim \hbar/2$) to describe the particles quasi-existing in our realistic world, whose studies may allow us to go beyond the quantum physics and arrive at another type of boundary of our realistic world, that is from the existent to the non-existent. We propose to investigate the quasi-existent particles by studying the singly heavy baryons through various imperfect symmetries among them. We obtain an incidental conjecture that the lifetime and width may be quantized for the particles existing in our realistic world as
\begin{equation*}M\tau=n\hbar/2~~~{\rm{and}}~~~M/\Gamma=n/2\,,~~~n=1,2,3\cdots\,.\end{equation*}
\end{abstract}
\maketitle
\pagenumbering{arabic}
%

%=====================================================================================
%=====================================================================================
\section{What do we learn from the hadron physics}
\label{sec:hadron}
%=====================================================================================
%=====================================================================================

The book of Chuang Tzu is an ancient and important Chinese spiritual text, where the author wrote:``Take a foot long stick and remove half every day. In ten thousands years it will not run out.'' This relates to the old philosophical question:``Is matter infinitely divisible?'' Although this question has not yet been fully answered, we have gained much knowledge about our microscopic world, {\it e.g.}, the divisibility stops at quarks and leptons according to the Standard Model, while some scientists believe that these elementary particles also have a structure~\cite{pdg}. A similar question is that:``Can the lifetime of an unstable particle be infinitely short?'' This is a philosophical question that has not yet been well investigated scientifically. The past decades witnessed the golden era of hadron physics, which gives us a good opportunity to investigate this question by studying the physics that happens in a transient period of time. Two of the major developments are about the singly heavy baryons and the exotic hadrons. Let us briefly introduce them as follows, and we refer to the recent review~\cite{Chen:2022asf} for their detailed discussions.

The electromagnetic interaction holds the atomic nucleus and electrons together inside an atom, and it leads to the fine/hyperfine structure of the light spectrum. The strong interaction among quarks and gluons is similar to the electromagnetic interaction in some aspects, and it can also lead to the fine structure of the hadron spectrum. An ideal platform to study this is the singly heavy baryon containing one
valence heavy quark ($charm/bottom$) and two valence light quarks ($up/down/strange$), where the light quarks and gluons circle around the nearly static heavy quark, so the whole system behaves as the QCD analogue of the atom bound by the electromagnetic interaction~\cite{Copley:1979wj,Korner:1994nh,Manohar:2000dt,Klempt:2009pi,Chen:2016spr}.

\begin{figure}[hbtp]
\begin{center}
\subfigure[]{\includegraphics[width=0.22\textwidth]{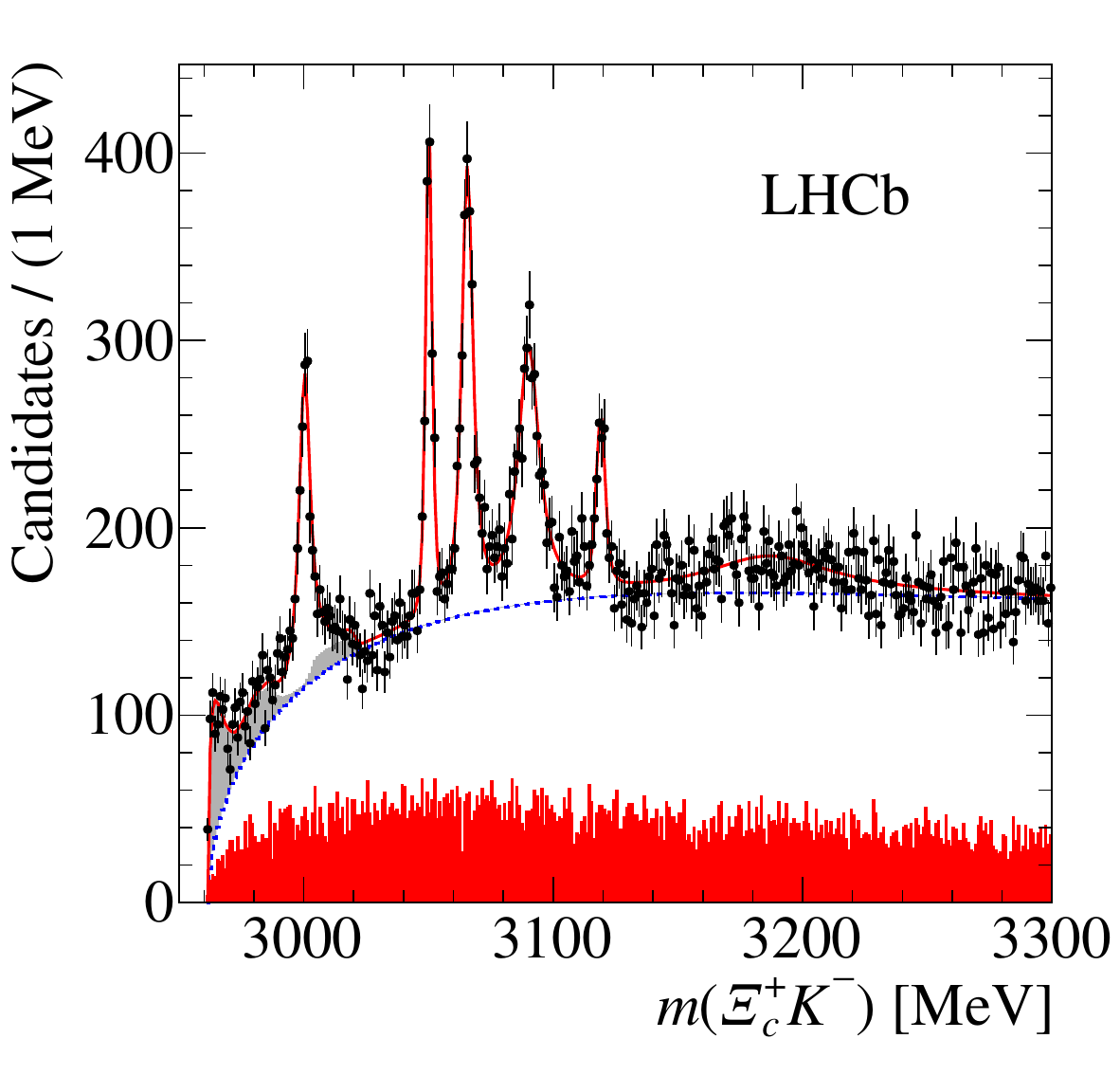}}
~~~
\subfigure[]{\includegraphics[width=0.22\textwidth]{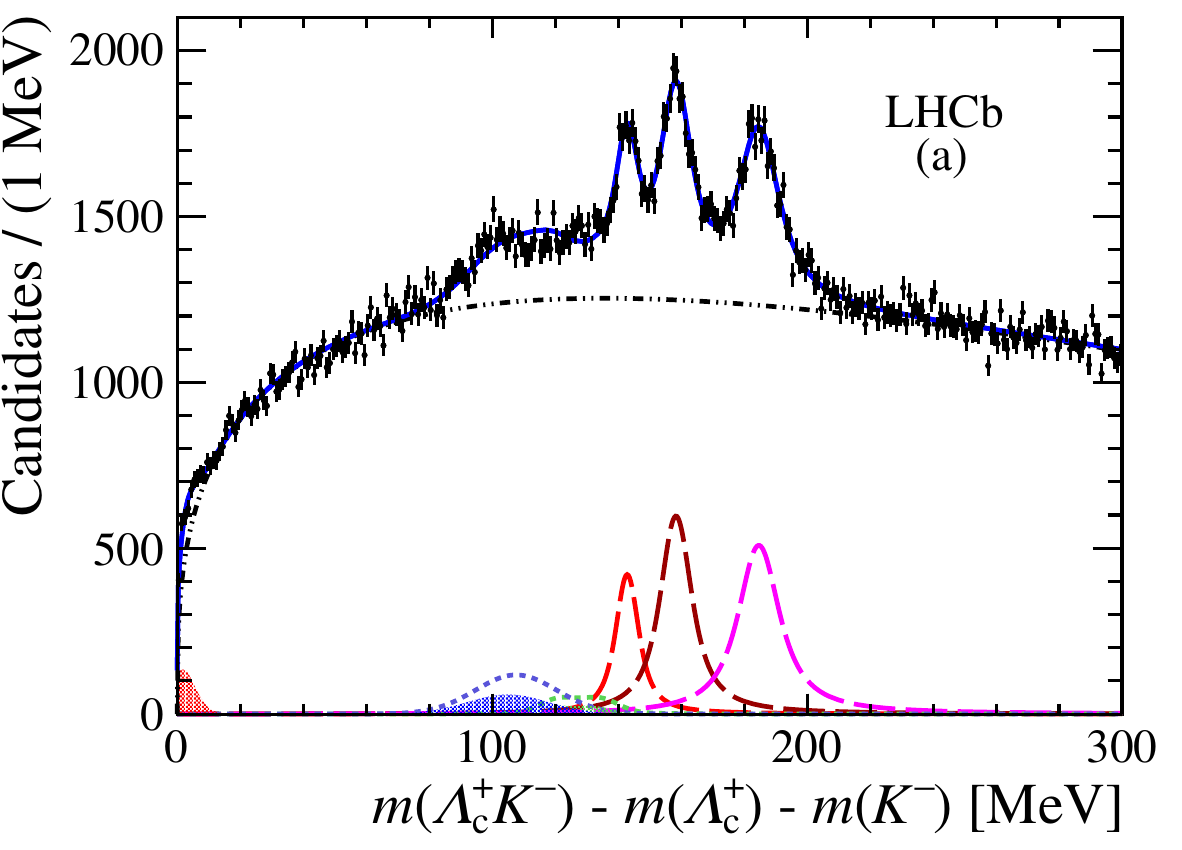}}
\subfigure[]{\includegraphics[width=0.4\textwidth]{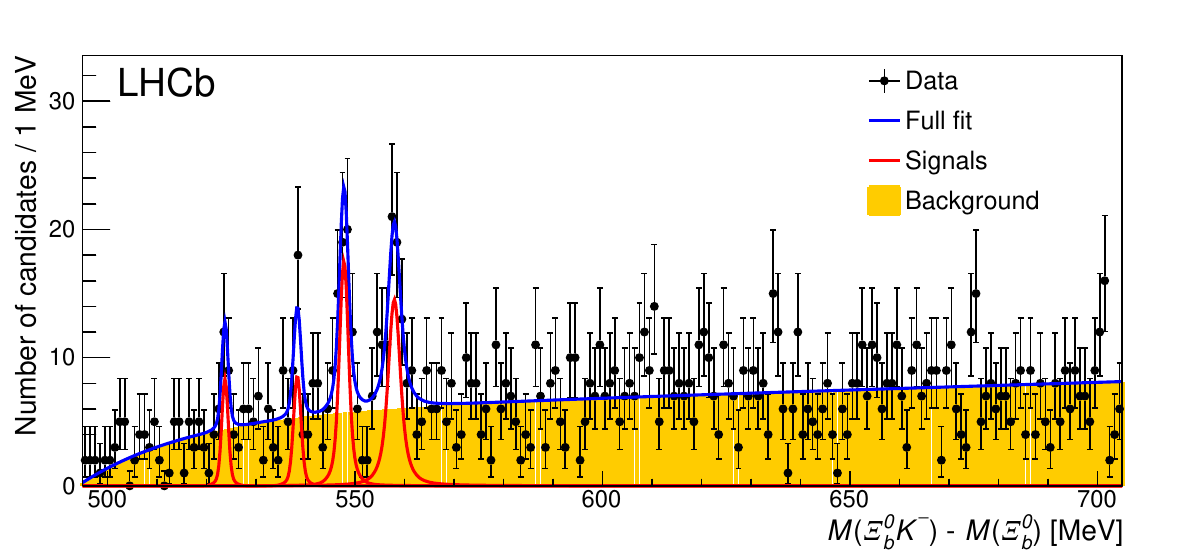}}
\end{center}
\caption{(a) The $\Xi^+_c K^-$ invariant mass spectrum measured by LHCb in 2017 and taken from Ref.~\cite{LHCb:2017uwr}. (b) Distribution of the invariant mass difference $\Delta m = m(\Lambda^+_c K^-) - m(\Lambda^+_c) - m(K^-)$, measured by LHCb in 2020 and taken from Ref.~\cite{LHCb:2020iby}. (c) Distribution of the invariant mass difference $\Delta m^\prime = m(\Xi_b^0 K^-) - m(\Xi_b^0)$, measured by LHCb in 2020 and taken from Ref.~\cite{LHCb:2020tqd}. These fine structures indicate the rich internal structures of the excited $\Omega_c/\Xi_c/\Omega_b$ baryons.}
\label{fig:heavybaryon}
\end{figure}

Important progresses have been made in this field in the past five years. In 2017 the LHCb collaboration discovered as many as five excited $\Omega_c^0$ baryons in the $\Xi_c^+ K^-$ mass spectrum~\cite{LHCb:2017uwr}, as shown in Fig.~\ref{fig:heavybaryon}(a):
\begin{equation*}
\Omega_c(3000)^0 , \, \Omega_c(3050)^0 , \, \Omega_c(3066)^0 , \, \Omega_c(3090)^0 , \, \Omega_c(3119)^0 .
\end{equation*}
In 2020 they discovered three excited $\Xi_c^0$ baryons in the $\Lambda^+_c K^-$ mass spectrum~\cite{LHCb:2020iby}, as shown in Fig.~\ref{fig:heavybaryon}(b):
\begin{equation*}
\Xi_c(2923)^0 \, , \, \Xi_c(2939)^0 \, , \, \Xi_c(2965)^0 \, .
\end{equation*}
Also in 2020 they discovered four excited $\Omega_b^-$ baryons in the $\Xi_b^0 K^-$ mass spectrum~\cite{LHCb:2020tqd}, as shown in Fig.~\ref{fig:heavybaryon}(c):
\begin{equation*}
\Omega_b(6316)^- \, , \, \Omega_b(6330)^- \, , \, \Omega_b(6340)^- \, , \, \Omega_b(6350)^- \, .
\end{equation*}
These excited $\Omega_c/\Xi_c/\Omega_b$ baryons are good candidates for the $1P$-wave singly heavy baryons. Their fine structures indicate the rich internal structures of hadrons, which are caused directly by the strong interaction.

An atomic nucleus is made of protons and neutrons, which are themselves composed of quarks and gluons. In the past century a huge number of hadrons were observed in particle experiments, which are similar to the proton and neutron. One naturally conjecture that these hadrons are capable of forming some subatomic particles that are similar to the atomic nucleus~\cite{Weinberg:1965zz,Voloshin:1976ap}. Nowadays we call these subatomic particles ``hadronic molecules''. Since the discovery of the $X(3872)/\chi_{c1}(3872)$ by Belle in 2003~\cite{Belle:2003nnu}, there have been significant progresses in this field and many candidates for the hadronic molecules have been observed in experiments. Note that some of these candidates can also be explained as the compact multiquark states that are still hadrons but not hadronic molecules~\cite{Jaffe:1976ig,Maiani:2004vq}.

\begin{figure}[hbtp]
\begin{center}
\subfigure[]{\includegraphics[width=0.22\textwidth]{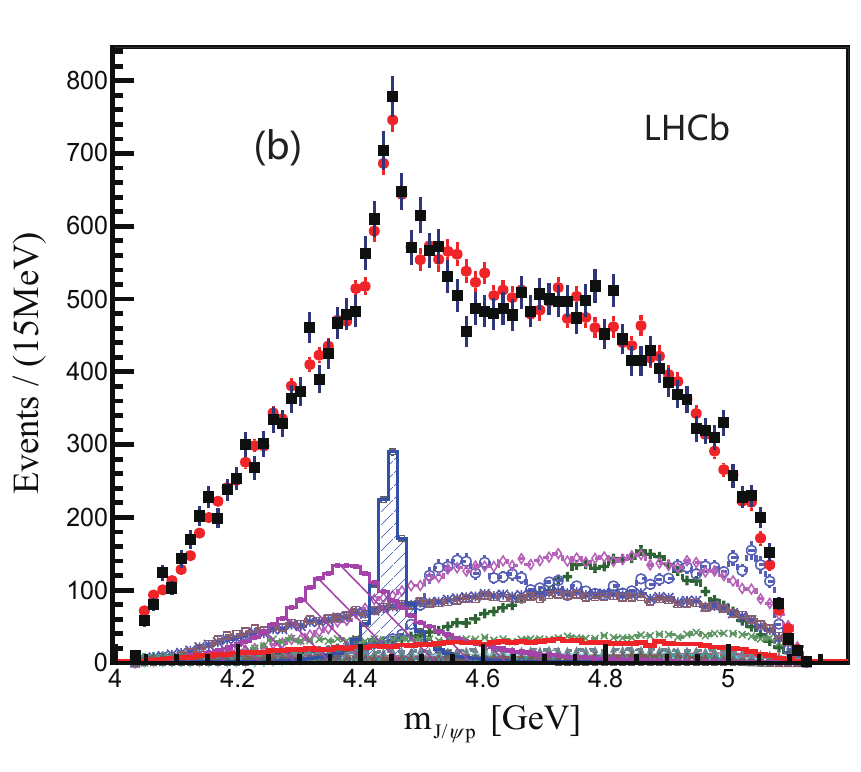}}
~~~
\subfigure[]{\includegraphics[width=0.22\textwidth]{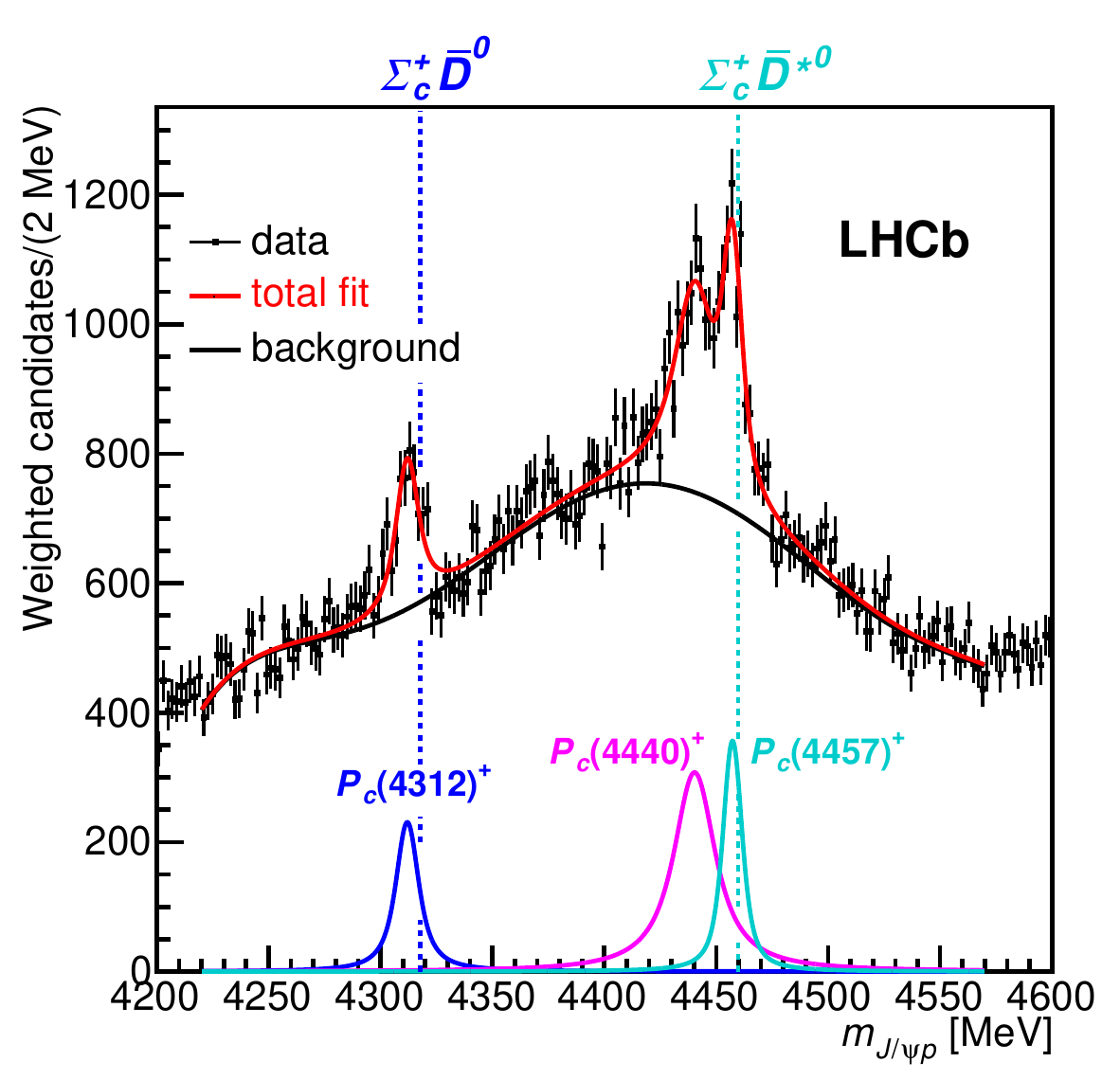}}
\end{center}
\caption{The $J/\psi p$ invariant mass spectrum of the $\Lambda_b^0 \to J/\psi K^- p$ decay, measured by LHCb (a) in 2015~\cite{LHCb:2015yax} and (b) in 2019~\cite{LHCb:2019kea}. The purple and blue solid curves in the left panel describe the $P_c(4380)^+$ and $P_c(4450)^+$, respectively. The latter $P_c(4450)^+$ is further separated into two substructures, $P_c(4440)^+$ and $P_c(4457)^+$, in the right panel. Taken from Refs.~\cite{LHCb:2015yax,LHCb:2019kea}.}
\label{fig:pc}
\end{figure}

As shown in Fig.~\ref{fig:pc}, in the LHCb experiments performed in 2015~\cite{LHCb:2015yax} and 2019~\cite{LHCb:2019kea}, the famous hidden-charm pentaquark states,
\begin{equation*}
P_c(4312)^+ \, , \, P_c(4380)^+ \, , \, P_c(4440)^+ \, , \, P_c(4457)^+ \, ,
\end{equation*}
were discovered; besides, in the LHCb experiments performed in 2020~\cite{LHCb:2020jpq} and 2022~\cite{Collaboration:2022boa}, the hidden-charm pentaquark states with strangeness,
\begin{equation*}
P_{cs}(4338)^0 \, , \, P_{cs}(4459)^0 \, ,
\end{equation*}
were discovered. These $P_c/P_{cs}$ structures are just below their corresponding meson-baryon thresholds, and their discoveries strongly support the existence of hadronic molecules~\cite{Wu:2010jy,Yang:2011wz,Wang:2011rga,Karliner:2015ina,Chen:2016qju}. Moreover, it was proposed in Ref.~\cite{Chen:2021xlu} that the residue strong interaction is capable of forming the covalent hadronic molecules through the shared light quarks, similar to the chemical molecules formed by the residue electromagnetic interaction through the shared electrons.

%
%=====================================================================================
%=====================================================================================
\section{Imaginable worlds}
\label{sec:imagine}
%=====================================================================================
%=====================================================================================
%

\begin{figure*}[hbtp]
\begin{center}
\includegraphics[width=0.84\textwidth]{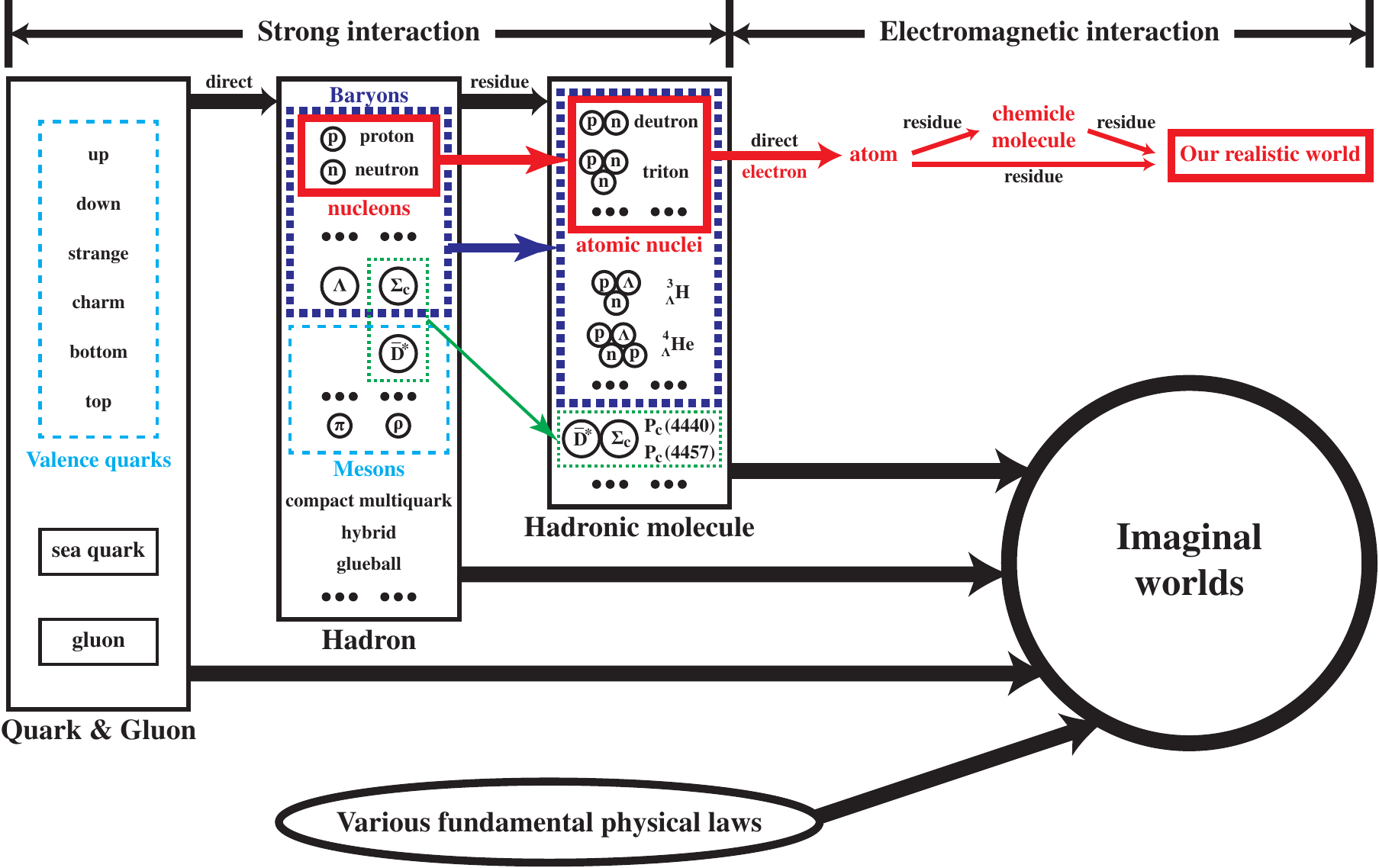}
\end{center}
\caption{A picture of the world from the viewpoint of hadron physics.}
\label{fig:world}
\end{figure*}

Based on the recent developments of the hadron physics, let us imagine a world without the weak interaction. To start with, we assume that there is only a small net number of the $strange$, $charm$, and $bottom$ quarks. In our realistic world most of the hadrons are not stable, while in this imaginal world some of them become stable besides the proton and neutron, such as
\begin{equation*}
\Lambda^0 \, , \, \Lambda_c^+ \, , \, \Lambda_b^+ \, \cdots
\end{equation*}
As depicted in Fig.~\ref{fig:world}, protons and neutrons compose various atomic nuclei, and then compose various atoms and chemical molecules together with electrons, which further compose our realistic world. After turning off the weak interaction, there exist more stable hadrons in the imaginal world, which are capable of forming more hadronic molecules besides the atomic nuclei. Accordingly, the imaginal world can be more complicated.

Let us further assume that the net number of the $strange$ quarks is larger than that of the $up$ and $down$ quarks, while the net number of the $charm$ and $bottom$ quarks is still small. In this imaginal world the stable hadrons are
\begin{equation*}
\Lambda^0 \, , \, \Xi^{0,-} \, , \, \Xi_c^{+,0} \, , \, \Xi_b^{0,-} \, \cdots
\end{equation*}
while the lower-lying proton and neutron are limited by the reactions:
\begin{equation}
p + \Xi^{-} \to 2 \Lambda^0 ~~~{\rm and}~~~ n + \Xi^{0} \to 2 \Lambda^0 \, .
\end{equation}
Hence, there can be various worlds formed by various fundamental physical laws with certain fundamental parameters, and the world can become much different as long as its fundamental physical laws or its fundamental parameters change a little. Especially, our realistic world is so ingenious that the free neutrons decay through the weak interaction, but the neutrons bound inside many atomic nuclei do not.

To get a feeling of the above imaginable worlds, we show several time scales as follows:
\begin{itemize}

\item The lifetime of our universe is $1.38 \times 10^{10}$~years~$\sim 10^{+17}$~s.

\item The lifetime of the neutron is $877.75$~s~$\sim 10^{+3}$~s.

\item The lifetimes of the $\Lambda^0$ and $\Xi^{0,-}$ hyperons are all at the order of $10^{-10}$~s.

\item The time needed to form a hadron is less than $1$~GeV$^{-1}$~$\sim 10^{-24}$~s.

\end{itemize}
We compare the lifetime of the $\Lambda^0$ hyperon with its forming time, and the ratio is about $10^{+14}$. We compare the lifetime of our universe with that of the neutron as its fundamental element, and the ratio is also about $10^{+14}$. Hence, the imaginable worlds, with a small number of the $p/n/\Lambda^0$ or $\Lambda^0/\Xi^{0}/\Xi^{-}$ baryons, may exist in a very short time, {\it e.g.}, $10^{-15}$~s. Besides these imaginable worlds made of hadrons, there might be the imaginable worlds directly made of quarks and gluons, which might exist in a time scale much shorter than that needed to form a hadron.

%
%=====================================================================================
%=====================================================================================
\section{Existing or not-existing}
\label{sec:existing}
%=====================================================================================
%=====================================================================================
%

In our realistic world most of the hadrons and hadronic molecules are not stable partly due to the weak interaction. Some of them have so transient lifetimes that they may not even be formed. Then are these non-existent particles still capable of affecting our realistic world? Since we can not directly observe them in experiments, we build a simple toy model towards an answer to this question:
\begin{itemize}

\item $A$ is a stable particle.

\item $B_1$ and $B_2$ both decay to $A$. The width of $B_1$ to $A$ is limited, while the width of $B_2$ to $A$ is so large that $B_2$ does not exist.

\item $C_1$ and $C_2$ both decay to $B_1$. The width of $C_1$ to $B_1$ is limited, while the width of $C_2$ to $B_1$ is so large that $C_2$ does not exist.

\item $C_3$ and $C_4$ both decay to $B_2$. The width of $C_3$ to $B_2$ is limited, while the width of $C_4$ to $B_2$ is so large that $C_4$ does not exist.

\item The widths of $C_{1\cdots4}$ to $A$ are all negligible.

\end{itemize}
Let us investigate two possibilities as follows.

\begin{figure}[hbtp]
\begin{center}
\subfigure[]{\includegraphics[width=0.21\textwidth]{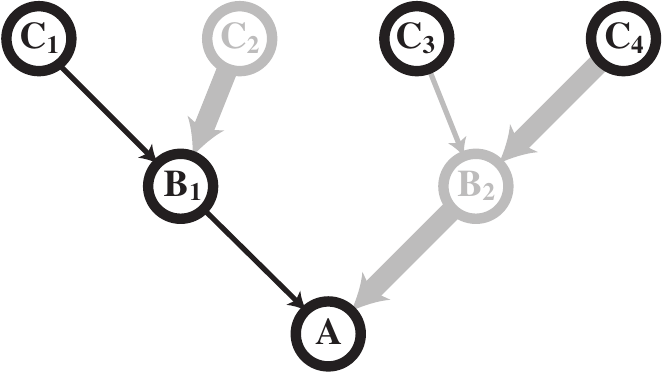}}
~~~~~
\subfigure[]{\includegraphics[width=0.21\textwidth]{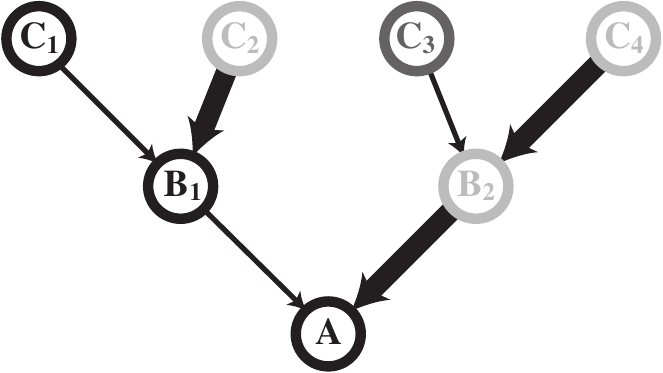}}
\end{center}
\caption{A simple toy model based on the assumptions: (a) the non-existent particles do not affect our realistic world, and (b) the non-existent particles can still affect our realistic world.}
\label{fig:existing}
\end{figure}

Firstly, we assume that the non-existent particles do not affect our realistic world. As depicted in Fig.~\ref{fig:existing}(a), we derive:
\begin{itemize}

\item $A$ is a stable particle, so it exists. The widths of $B_1$ and $C_1$ are both limited, so they also exist. The widths of $B_2$ and $C_2$ are both too large, so they do not exist.

\item Under the assumption that the non-existent particle $B_2$ does not affect our realistic world, the particles $C_3$ and $C_4$ are both stable, so they exist.

\end{itemize}
The last item is not so reasonable that the present assumption is problematic.

Secondly, we assume that the non-existent particles can still affect our realistic world. As depicted in Fig.~\ref{fig:existing}(b), we derive:
\begin{itemize}

\item $A$ is a stable particle, so it exists. The widths of $B_1$ and $C_1$ are both limited, so they also exist. The widths of $B_2$ and $C_2$ are both too large, so they do not exist.

\item Under the assumption that the non-existent particle $B_2$ can still affect our realistic world, the width of $C_4$ to $B_2$ is too large, so $C_4$ does not exist.

\item Under the assumption that the non-existent particle $B_2$ can still affect our realistic world, $C_3$ may exist since the width of $C_3$ to $B_2$ is limited, while it does not exist if the width of the decay process $C_3 \to B_2 \to A$ turns out to be too large.

\end{itemize}
These analyses are more reasonable than the previous analyses, so the present assumption is more reasonable than the previous one. We may further derive from this assumption that the higher-lying particles are generally more unstable.

Based on the above analyses, we conjecture that there are hadrons and hadronic molecules that are too unstable to exist, but they can still affect our realistic world. We may further conjecture that the imaginal worlds, made of these non-existent particles, do not exist but still weakly affect our realistic world. Note that some new physics particles may behave similarly to the intermediate particle $B_2$, given that their widths are too large for them to be observed in our realistic world.

%
%=====================================================================================
%=====================================================================================
\section{Beyond the quantum physics}
\label{sec:quantum}
%=====================================================================================
%=====================================================================================
%

In this section we further discuss what kinds of particles exist, and so can be observed, in our realistic world. Note that the concept of ``exist'' has been widely discussed in philosophy, while in this study we simply treat it the same as ``observable'' or ``capable of being observed''.

Undoubtedly, the answer closely relates to the lifetimes of particles. All the stable hadron (the proton) and hadronic molecules (some of the atomic nuclei) have been well established. Some of the unstable hadrons and hadronic molecules can be observed in experiments, given that their widths are not too large. On the contrary, some of the unstable hadrons and hadronic molecules have so transient lifetimes that they may not even be formed, not to mention being observed in experiments.

However, the answer not only relates to the lifetimes, for examples:
\begin{itemize}

\item The widths of the $W$ and $Z$ bosons are $2.085 \pm 0.042$~GeV and $2.4952 \pm 0.0023$~GeV, respectively. The width of the $top$ quark is $1.42^{+0.19}_{-0.15}$~GeV, that is also larger than 1~GeV. Although these three particles are all very broad, they have been well established in experiments. For comparisons, their masses are $80.379 \pm 0.012$~GeV, $91.1876 \pm 0.0021$~GeV, and $172.76 \pm 0.30$~GeV, respectively.

\item However, all the hadrons well observed (or at least not badly observed) in experiments have the widths smaller than 1~GeV. As two of the broadest hadrons observed so far, the $\pi_2(2100)$ meson and the $N(2600)$ baryon have the widths $625 \pm 50$~MeV and $500 - 800$~MeV, respectively. For comparisons, their masses are $2090 \pm 29$~MeV and $2550 - 2750$~MeV, respectively.

\item The pole parameter of the $f_0(500)$ meson is estimated in the PDG2020~\cite{pdg} to be $(400-550)-i(200-350)$~MeV, so its mass and width are
\begin{eqnarray}
M_{f_0(500)} &=& 400 - 550 {~\rm MeV} \, ,
\\ \nonumber \Gamma_{f_0(500)} &=& 400 - 700 {~\rm MeV} \, .
\end{eqnarray}
Besides, the pole parameter of the $K_0^*(700)$ meson is estimated to be $(630 - 730) -i (260 - 340)$~MeV~\cite{pdg}, so its mass and width are
\begin{eqnarray}
M_{K_0^*(700)} &=& 630 - 730 {~\rm MeV} \, ,
\\ \nonumber \Gamma_{K_0^*(700)} &=& 520 - 680 {~\rm MeV} \, .
\end{eqnarray}
The establishments of these two mesons are probably the most difficult among all the hadrons~\cite{Pelaez:2015qba}.

\end{itemize}
Therefore, the answer seems to relate to not only the lifetimes but also the masses of particles.

Generally speaking, the particle with a larger ratio
\begin{equation}
R \equiv M/\Gamma \sim M \tau / \hbar \, ,
\end{equation}
can be more easily observed, while that with a smaller ratio becomes more difficult to be observed. Here $M$, $\Gamma$, and $\tau$ are the mass, width, and lifetime of the particle, respectively.

Especially, the $f_0(500)$ and $K_0^*(700)$ mesons both satisfy $R \equiv M/\Gamma \sim 1$. We use $\Delta E$ and $\Delta t$ to denote the uncertainties of $M$ and $\tau$, respectively. Given that the mass $M$ and the lifetime $\tau$ are two well-defined positive-definite parameters, their uncertainties should be smaller than themselves:
\begin{eqnarray}
\Delta E &\lesssim& M \, ,
\\ \Delta t &\lesssim& \tau \, ,
\end{eqnarray}
so the $f_0(500)$ and $K_0^*(700)$ mesons further satisfy
\begin{equation}
\Delta E \Delta t \lesssim M \tau \sim M \hbar / \Gamma = R \hbar \sim \hbar \, .
\end{equation}
This indicates that the uncertainty principle $\Delta E \Delta t \geqslant \hbar/2$ may play an important role in determining ``what kinds of particles exist, and so can be observed, in our realistic world''.

Let us consider an unstable particle with $R \equiv M/\Gamma < 1/2$ ($M \tau < \hbar/2$). This particle does not exist in our realistic world probably, since it does not obey the uncertainty principle:
\begin{equation}
\Delta E \Delta t \lesssim M \tau \sim M \hbar / \Gamma = R \hbar < \hbar/2 \, .
\end{equation}
This follows the argument that the particles existing in our world should satisfy some of its fundamental physical laws such as the quantum physics, while those not-existing do not need to. We shall further discuss this at the end of this paper.

Another example is the sea quarks confined inside hadrons, which do not satisfy the uncertainty principle $\Delta p \Delta x \geqslant \hbar/2$, and correspondingly, the free quarks do not exist in our realistic world. This triggers the question on the definition of ``free'': does the free quark still not exist in the time scale much shorter than that needed to form a hadron, and correspondingly, does the free electric charge still exist in the time scale larger than the lifetime of our universe?

\begin{figure}[hbtp]
\begin{center}
\includegraphics[width=0.48\textwidth]{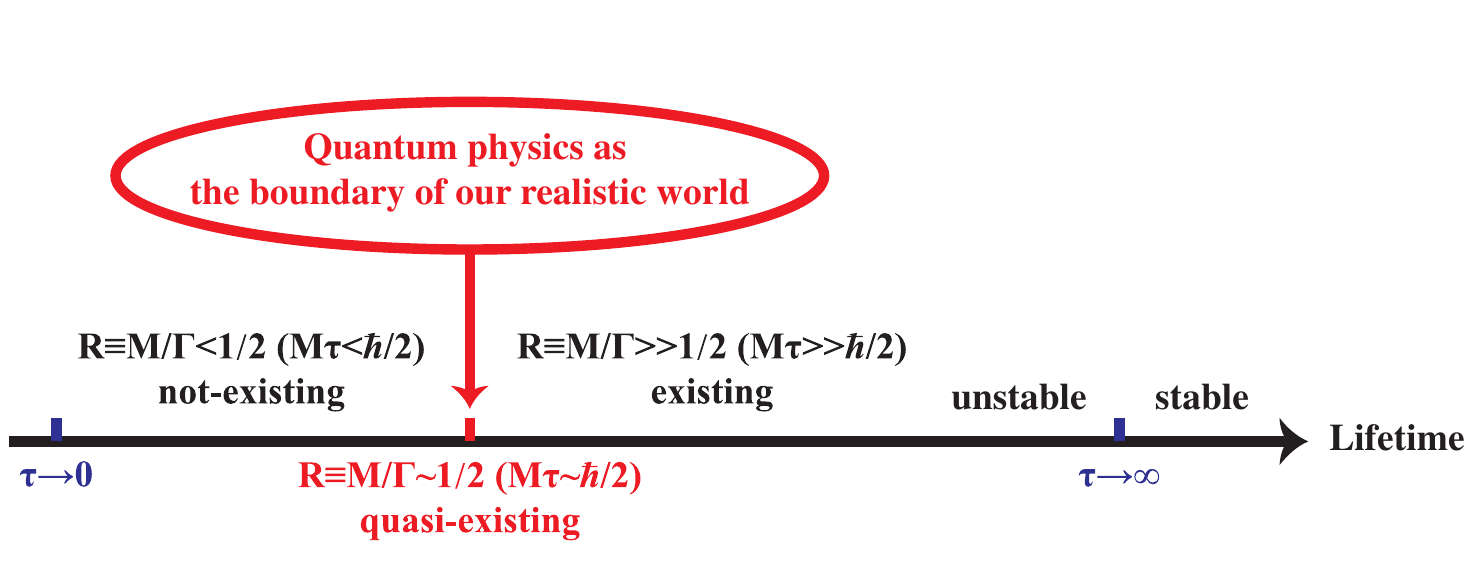}
\end{center}
\caption{Another type of boundary of our realistic world, that is from the existent to the non-existent. The ratio $R$ is defined as $R \equiv M/\Gamma \sim M \tau / \hbar$, where $M$, $\Gamma$, and $\tau$ represent the mass, width, and lifetime of the unstable particle, respectively.}
\label{fig:beyond}
\end{figure}

Based on the above analyses, we conjecture that the ratios $R \equiv M/\Gamma \gg 1/2$ ($M \tau \gg \hbar/2$) and $R \equiv M/\Gamma < 1/2$ ($M \tau < \hbar/2$) can be used to describe the particles existing and not-existing in our realistic world, respectively. Besides, the ratio $R \equiv M/\Gamma \sim 1/2$ ($M \tau \sim \hbar/2$) can be used to describe the particles quasi-existing in our realistic world. These quasi-existent particles are of particular interest, since they may not well obey the quantum physics but they may still be observed in experiments. As depicted in Fig.~\ref{fig:beyond}, studying these quasi-existent particles may allow us to go beyond the quantum physics and arrive at another type of boundary of our realistic world, that is from the existent to the non-existent.

%
%=====================================================================================
%=====================================================================================
\section{The singly heavy baryon system as a possible platform}
\label{sec:baryon}
%=====================================================================================
%=====================================================================================
%

As discussed in the previous section, the quasi-existent particles satisfying $R \equiv M/\Gamma \sim 1/2$ ($M \tau \sim \hbar/2$) are of particular interest, since they may not well obey the quantum physics but they may still be observed in experiments. Especially, the $f_0(500)$ and $K_0^*(700)$ mesons both satisfy $R \equiv M/\Gamma \sim 1$, so they are worthy of further investigations. However, as isolated cases, their studies are probably not enough. Instead, we propose to study the singly heavy baryon system, which may serve as an even better platform due to various imperfect symmetries among the many singly heavy baryons.

As discussed in Sec.~\ref{sec:hadron}, the singly heavy baryon contains one valence heavy quark ($charm/bottom$) and two valence light quarks ($up/down/strange$), where the light quarks and gluons circle around the nearly static heavy quark. This system behaves as the QCD analogue of the atom bound by the electromagnetic interaction, but its internal structure is much more complicated.

\begin{figure}[hbtp]
\begin{center}
\includegraphics[width=0.25\textwidth]{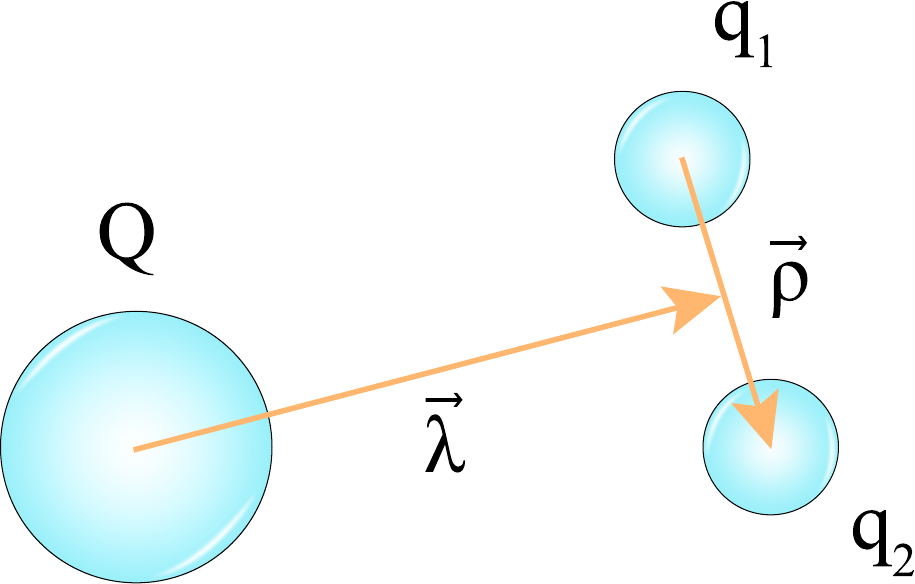}
\end{center}
\caption{Jacobi coordinates $\vec \lambda$ and $\vec \rho$ for the singly heavy baryon system.}
\label{fig:Jacobi}
\end{figure}

\begin{figure*}[hbtp]
\begin{center}
\includegraphics[width=0.8\textwidth]{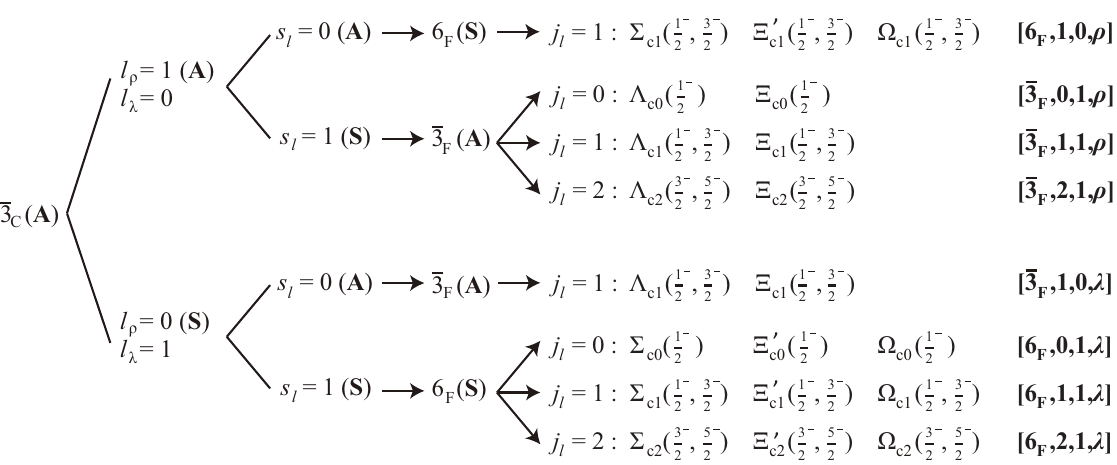}
\end{center}
\caption{Categorization of the $1P$-wave singly charmed baryons into eight multiplets within the framework of heavy quark effective theory.}
\label{fig:categorization}
\end{figure*}

As depicted in Fig.~\ref{fig:Jacobi}, we focus on the $Qq_1q_2$ system, with $Q$ the heavy quark and $q_{1,2}$ the two light quarks:
\begin{itemize}

\item Its total orbital angular momentum is
\begin{equation}
L = l_\rho \otimes l_\lambda \, ,
\end{equation}
where $l_\rho$ is the orbital angular momentum between the two light quarks, and $l_\lambda$ is the orbital angular momentum between the heavy quark and the two-light-quark system.

\item Its total spin angular momentum is
\begin{equation}
S = {s}_{q_1} \otimes {s}_{q_2} \otimes {s}_Q = s_l \otimes {s}_Q \, ,
\end{equation}
where $s_{q_{1,2}}$ and $s_Q$ are the light and heavy quark spin angular momenta, and $s_l = {s}_{q_1} \otimes {s}_{q_2}$ is the total spin angular momentum of the two light quarks.

\item In the heavy quark limit $m_Q \to \infty$, its properties are mainly determined by the light degrees of freedom. Accordingly, its total angular momentum
\begin{equation}
{J}=\left(l_\rho \otimes l_\lambda\right)_L \otimes \left({s}_{q_1} \otimes {s}_{q_2} \otimes {s}_Q\right)_{S} \, ,
\end{equation}
can be written as
\begin{equation}
{J}= \left(\left(l_\rho \otimes l_\lambda\right)_L \otimes \left({s}_{q_1} \otimes {s}_{q_2}\right)_{s_l}\right)_{j_l} \otimes {s}_Q \, ,
\end{equation}
where $j_l = L \otimes s_l = l_\rho \otimes l_\lambda \otimes {s}_{q_1} \otimes {s}_{q_2}$ is the total angular momentum of the light component.

\end{itemize}

To well understand the singly heavy baryon system, we need to carefully examine the four internal symmetries between the two light quarks:
\begin{itemize}

\item The color structure of the two light quarks is antisymmetric ($\mathbf{\bar 3}_C$).

\item The flavor structure of the two light quarks is either antisymmetric or symmetric. Within the flavor $SU(3)$ symmetry, the antisymmetric structure is the $\mathbf{\bar 3}_F$ representation composed of the $\Lambda_Q/\Xi_Q$ baryons, and the symmetric structure is the $\mathbf{6}_F$ representation composed of the $\Sigma_Q/\Xi^\prime_Q/\Omega_Q$ baryons.

\item The spin structure of the two light quarks is either antisymmetric ($s_l = 0$) or symmetric ($s_l = 1$).

\item The orbital structure of the two light quarks is either antisymmetric ($l_\rho = 1/3/\cdots$) or symmetric ($l_\rho = 0/2/\cdots$).

\end{itemize}
According to the Pauli principle, the total wave function of the two light quarks is antisymmetric, based on which we can categorize the singly heavy baryons into several multiplets. According to the heavy quark effective theory~\cite{Grinstein:1990mj,Eichten:1989zv,Falk:1990yz}, two states can have the same quantum numbers $(j_l, s_l, l_\rho, l_\lambda)$ when $j_l \neq 0$, and they form a degenerate doublet with $J = j_l \otimes s_Q = j_l \pm 1/2$; there is only one state having the quantum numbers $(j_l, s_l, l_\rho, l_\lambda)$ when $j_l = 0$, and it forms a singlet itself with $J = j_l \otimes s_Q = 1/2$.

As shown in Fig.~\ref{fig:categorization}, we can categorize the $1P$-wave singly charmed baryons into eight multiplets, denoted as $[Flavor, j_l, s_l, \rho/\lambda]$. There can be as many as seven $\Lambda_c$ and seven $\Xi_c$ baryons, which form four multiplets of the $\mathbf{\bar 3}_F$ representation; there can also be seven $\Sigma_c$, seven $\Xi^\prime_c$, and seven $\Omega_c$ baryons, which form four multiplets of the $\mathbf{6}_F$ representation. As discussed in Sec.~\ref{sec:hadron}, their rich internal structure causes the fine structure of the strong interaction, which has been observed in recent LHCb experiments~\cite{LHCb:2017uwr,LHCb:2020iby,LHCb:2020tqd}.

In the past years important experimental progresses have been made in the field of singly heavy baryon system, and many candidates for the excited singly heavy baryons have been observed in experiments~\cite{pdg}. These observations have attracted much attention among the theorists, and various theoretical methods have been applied to study the singly heavy baryons~\cite{Chen:2022asf}. Especially, both the constitute quark model and the QCD sum rule method have been applied to systematically study their mass spectra and decay properties. The results obtained using these two methods are roughly consistent with each other, although not exactly the same, {\it e.g.}, see the constitute quark model calculations of Ref.~\cite{Liang:2020hbo} and the QCD sum rule calculations of Refs.~\cite{Yang:2020zrh,Yang:2021lce}.

Based on these two theoretical methods, some singly heavy baryons are calculated to have rather large widths, for examples:
\begin{itemize}

\item The width of the $\Omega_{b0}(1/2^-)$ baryon belonging to the $[\mathbf{6}_F,0,1,\lambda]$ singlet is calculated to be about 1.0~GeV in Ref.~\cite{Liang:2020hbo} through the constitute quark model. This width is calculated to be about 2.7~GeV in Ref.~\cite{Yang:2020zrh} through the QCD sum rule method. Note that the uncertainties of these two calculations are rather large, so these two results are still roughly consistent with each other. For comparison, the mass of this baryon is calculated to be about 6.3~GeV.

\item The width of the $\Omega_{c0}(1/2^-)$ baryon belonging to the $[\mathbf{6}_F,0,1,\lambda]$ singlet is calculated to be about 1.0~GeV in Ref.~\cite{Yang:2021lce} through the QCD sum rule method. For comparison, its mass is calculated to be about 3.0~GeV.

\item Especially, the widths of the $\Lambda_{c1}(1/2^-)$ and $\Xi_{c1}(1/2^-)$ baryons belonging to the $[\mathbf{\bar 3}_F,1,0,\lambda]$ doublet are calculated in Ref.~\cite{Yang:2022oog} through the QCD sum rule method to be about 2.4~GeV and 6.6~GeV, respectively. For comparisons, their masses are calculated to be about 2.7 and 2.9~GeV, respectively.

\end{itemize}
Besides, there can exist the $1D$- and $1F$-wave singly heavy baryons, etc. As higher-lying particles, their widths are generally larger than the $1P$-wave ones.

Therefore, there might exist some singly heavy baryons satisfying $R \equiv M/\Gamma \sim 1/2$. Moreover, there exist various imperfect symmetries in the singly heavy baryon system:
\begin{itemize}

\item The fine structure of the singly heavy baryons within the same multiplet. In the heavy quark limit the properties of a singly heavy baryon are mainly determined by its light degrees of freedom, so the baryons within the same multiplet are similar to some extent. For example, the $\Lambda_{c1}(1/2^-)$ and $\Lambda_{c1}(3/2^-)$ baryons belonging to the $[\mathbf{\bar 3}_F,1,0,\lambda]$ doublet are similar to each other to some extent.

\item The heavy quark flavor symmetry between the $charm$ and $bottom$ quarks. This symmetry indicates that the singly charmed baryons $\Lambda_c/\Xi_c/\Sigma_c/\Xi^\prime_c/\Omega_c$ and the singly bottom baryons $\Lambda_b/\Xi_b/\Sigma_b/\Xi^\prime_b/\Omega_b$ are respectively similar to each other. However, the heavy quark effective theory works better for the singly bottom baryons than for the singly charmed baryons, since the mass of the $bottom$ quark is significantly larger than that of the $charm$ quark.

\item The isospin $SU(2)$ symmetry between the $up$ and $down$ quarks as well as the flavor $SU(3)$ symmetry among the $up$, $down$, and $strange$ quarks. The former symmetry works quite well, while the latter still works but not so well. For examples, the former symmetry indicates that the singly bottom baryons $\Xi_b^0$ and $\Xi_b^-$ are quite similar to each other, and the latter indicates that the singly charmed baryons $\Sigma_c$, $\Xi^\prime_c$, and $\Omega_c$ are similar to some extent.

\end{itemize}

Although the singly heavy baryons with $R \equiv M/\Gamma \sim 1/2$ are rather difficult to be observed in experiments, their partner states due to the above imperfect symmetries may have a significantly larger ratio $R \equiv M/\Gamma$, and so become much easier to be observed. For example, there are five excited $\Omega_c^0$ baryons but only three excited $\Xi_c^0$ baryons observed in the LHCb experiments~\cite{LHCb:2017uwr,LHCb:2020iby}, as shown in Fig.~\ref{fig:heavybaryon}(a,b). Correspondingly, there can be two excited $\Xi_c^0$ baryons still missing, and they are expected to have significantly larger widths~\cite{Yang:2021lce}. The above imperfect symmetries can be applied to overcome some of the difficulties in studying these missing states.

There might exist some other symmetries and more fine/hyperfine structures in the singly heavy baryon system. These possibly-existing symmetries and the above imperfect symmetries make the singly heavy baryon system serve as a possible platform to study the quasi-existent particles with $R \equiv M/\Gamma \sim 1/2$ ($M \tau \sim \hbar/2$).

%
%=====================================================================================
%=====================================================================================
\section{Summary and Perspective}
\label{sec:summary}
%=====================================================================================
%=====================================================================================
%

In this paper we discuss what do we learn from the hadron physics after its quick developments in the past decades. The development on the singly heavy baryons indicates that there exists the fine structure of hadron spectrum caused by the direct strong interaction, and the development on the exotic hadrons indicates that the residue strong interaction is capable of forming the hadronic molecules. As depicted in Fig.~\ref{fig:world}, the direct and residue electromagnetic interactions form atoms and chemical molecules respectively, which further compose our realistic world. Similarly, the strong interaction may also be capable of forming some imaginable hadronic worlds, although these worlds might only exist in a very short time.

In our realistic world most of the hadrons and hadronic molecules are not stable. Some of them have so transient lifetimes that they may not even be formed. We discuss whether these non-existent particles are still capable of affecting our realistic world. Based on a simple toy model, we argue its answer to be possibly positive. Accordingly, we conjecture that there are hadrons and hadronic molecules that are too unstable to exist, but they can still affect our realistic world; there might be the imaginal worlds made of these non-existent particles, and these worlds do not exist neither but still weakly affect our realistic world.

We further discuss what kinds of particles exist, and so can be observed, in our realistic world. We argue that its answer possibly relates to not only the lifetimes but also the masses of particles, and moreover, the uncertainty principle may play an important role in answering this question. We conjecture that the ratios $R \equiv M/\Gamma \gg 1/2$ ($M \tau \gg \hbar/2$) and $R \equiv M/\Gamma < 1/2$ ($M \tau < \hbar/2$) can be used to describe the particles existing and not-existing in our realistic world, respectively. Here $M$, $\Gamma$, and $\tau$ are the masses, widths, and lifetimes of particles, respectively.

Actually, we can use a rather simple argument to show that the particle with $R \equiv M/\Gamma < 1/2$ ($M \tau < \hbar/2$) does not obey the uncertainty principle. Let us imagine an unstable particle with the mass $M$ and the lifetime $\tau \rightarrow 0$. The uncertainty principle demands
\begin{equation}
\Delta E \Delta t \geqslant {\hbar / 2 } \, ,
\end{equation}
where $\Delta E$ and $\Delta t$ are the uncertainties of $M$ and $\tau$, respectively. Given that the mass $M$ and the lifetime $\tau$ are two well-defined positive-definite parameters, their uncertainties should be smaller than themselves:
\begin{eqnarray}
\Delta E &\lesssim& M \, ,
\\ \Delta t &\lesssim& \tau \, .
\end{eqnarray}
Therefore, the uncertainty principle further demands
\begin{equation}
M \tau \gtrsim \hbar/2 \, ,
\end{equation}
and the particle with $R \equiv M/\Gamma < 1/2$ ($M \tau < \hbar/2$) does not obey the uncertainty principle.

The above analyses indicate that the quantum physics, which rules our microscopic world, may set a lower limit on the lifetime of the particle existing in our realistic world. To better understand this, we relate the Bohr's quantization condition for the angular momentum of the electron around the proton
\begin{equation}
L = m v r = n \hbar \, , ~~~ n=1,2,3\cdots \, ,
\end{equation}
to the condition of the circular standing wave
\begin{equation}
2 \pi r = n \lambda \, ,
\end{equation}
through the de Broglie relation $\lambda = h/p = h/(mv)$. Comparing with the two uncertainty principles $\Delta p \Delta x \geqslant \hbar/2$ and $\Delta E \Delta t \geqslant {\hbar / 2}$ as well as the correspondences
\begin{equation}
(p, \Delta p) \rightarrow (M, \Delta E) \, ,~ (r, \Delta x) \rightarrow (\tau, \Delta t) \, ,
\end{equation}
we conjecture that the lifetime may also be quantized:
\begin{equation}
M \tau = n \hbar \, , ~~~ n=1,2,3\cdots \, .
\end{equation}
Considering that the angular momentum quantum number can be either an integer or a half-integer actually, this quantization condition is finalized to be
\begin{equation}
M \tau = n \hbar/2 {~~~\rm and ~~~} M / \Gamma = n/2 \, , ~~~ n=1,2,3\cdots \, ,
\end{equation}
for the particles existing in our realistic world as the ``standing waves'' in the time axis~\footnote{This is a tentative conjecture; other forms such as $M \tau = n \hbar$ and $M / \Gamma = n$ ($n = 1, 2, 3, \cdots$) may also be plausible.}.

Besides the existent and non-existent particles, we propose to use the ratio $R \equiv M/\Gamma \sim 1/2$ ($M \tau \sim \hbar/2$) to describe the particles quasi-existing in our realistic world. These quasi-existent particles are of particular interest, since they may not well obey the quantum physics but they may still be observed in experiments. Their studies may allow us to go beyond the quantum physics and arrive at another type of boundary of our realistic world, that is from the existent to the non-existent. Especially, the $f_0(500)$ and $K_0^*(700)$ mesons both satisfy $R \equiv M/\Gamma \sim 1$, so they are worthy of further investigations. Moreover, we propose to study the singly heavy baryon system, which serves as a possible platform to study the quasi-existent particles due to various imperfect symmetries among the many singly heavy baryons.

Based on the above analyses, we further conjecture that some fundamental physical laws, such as the quantum physics, rule our realistic world simply because they are the boundaries stabilizing our realistic world (or blocking our realistic world from some other worlds); the particles existing in our realistic world should satisfy these fundamental physical laws, while those not-existing do not need to. This argument has already been stated in Sec.~\ref{sec:quantum}. Therefore, we may meet some new physics when investigating the particles not-existing in our realistic world, and we might also meet some other new physics when investigating the particles quasi-existing in our realistic world, {\it e.g.}, the non-existent particle $B_2$ depicted in Fig.~\ref{fig:existing}(b) may be investigated through the decay process $C_3 \to B_2 \to A$, and the possibly-existing quasi-existent singly heavy baryons may be investigated through various imperfect symmetries among the many singly heavy baryons.

Besides our realistic world, there can be various worlds formed by various fundamental physical laws with certain fundamental parameters, and the world can become much different as long as its fundamental physical laws or its fundamental parameters change a little. Especially, there may exist the imaginal hadronic worlds formed by the strong interaction, whose time and space scales are much smaller than those of our realistic world. If there exist some civilizations in these mini-worlds, there might be a tiny chance that these civilizations are willing and able to send out some of their information, and at the same time we are able to receive these information.

To end this paper, let us imagine that someday in the future our human beings finally exhaust all the resources of our universe, but still we can not break through its boundary. What we can do at that time might be exchanging our belongings to some external world for the necessary resources, and the most valuable belonging of ours is probably the knowledge of our universe. On the contrary, we might also serve as the external world, and exchange our resources to the civilizations from some mini-worlds for the knowledge of their universes.

%
%=====================================================================================
%=====================================================================================
%=====================================================================================
\section*{Acknowledgments}
%=====================================================================================
%=====================================================================================
%=====================================================================================
%

This project is supported by
the National Natural Science Foundation of China under Grant No.~12075019,
the Jiangsu Provincial Double-Innovation Program under Grant No.~JSSCRC2021488,
and
the Fundamental Research Funds for the Central Universities.

\bibliographystyle{elsarticle-num}
\bibliography{ref}

\end{document}